\documentclass[reprint,
superscriptaddress,
nofootinbib,
amsmath,amssymb,
aps,
]{revtex4-2}
\usepackage[colorlinks=true,linkcolor=blue,citecolor=blue,urlcolor=
blue]{hyperref}

\usepackage[utf8]{inputenc}
\usepackage{float}
\usepackage{graphicx}% Include figure files
\usepackage{dcolumn}% Align table columns on decimal point
\usepackage{bm}% bold math
\usepackage{amsmath}
\usepackage{tensor}
\usepackage[dvipsnames]{xcolor}
\usepackage{subcaption}
\usepackage{lipsum}
\usepackage{mwe}

\usepackage{graphicx}
\usepackage{subcaption}
\begin{document}
	
	\preprint{APS/123-QED}

    \title{Dominance of Electric Fields in the Charge Splitting of Elliptic Flow }
    %\title{Temporal Evolution of Electromagnetic Fields in Low-Energy Au+Au Collisions with Baryon Stopping}
    \author{Ankit Kumar Panda}
    \email{ankitkumarpanda932@gmail.com}
    %\author{Partha Bagchi}
    %\email{parphy@niser.ac.in}
     %\author{Hiranmaya Mishra}
     %\email{hiranmaya@niser.ac.in}
    %\author{Victor Roy}
    %\email{victor@niser.ac.in}

    \affiliation{School of Physical Sciences, Indian Institute of Technology Goa, Ponda-403401, Goa, India}

	\date{\today}% It is always \today, today,
	
	\begin{abstract}

   In this study, we investigate the impact of electromagnetic fields, highlighting the dominant effect of electric fields on the splitting of elliptic flow, \( \Delta v_2 \) with transverse momentum ($p_T$).
The velocity and temperature profiles of quark-gluon plasma (QGP) is described through thermal model calculations. The electromagnetic field evolution is however determined from the solutions of Maxwell's equations, assuming constant electric and chiral conductivities. We find that the slower decay of the electric fields compared to the magnetic fields makes its impact on the splitting of the elliptic flow more dominant.
We further estimated that the maximum value of \( |\langle eF \rangle| \), evaluated by averaging the field values over all spatial points on the hypersurface and across all field components, is approximately \( (0.010003  \pm 0.000195) \, m_{\pi}^2 \) for \( \sqrt{s_{\text{NN}}} = 7.7 \, \text{GeV} \), which could describe the splitting of elliptic flow data within the current experimental uncertainty reasonably well.

	\end{abstract}

    \maketitle
    \section{Introduction}

In high-energy off-central heavy-ion collisions, it is well established that substantial electromagnetic fields are generated in the initial stages of heavy-ion collisions alongside the strongly coupled~\cite{Romatschke:2007mq,Song:2007ux,Kovtun:2004de} quark-gluon plasma (QGP) medium. These fields thus generated can reach magnitudes of up to $10^{18}$ to $10^{19}$ Gauss at the highest center-of-mass energies observed at RHIC and LHC~\cite{Skokov:2009qp,Voronyuk:2011jd,Bzdak:2011yy,Deng:2012pc,Kharzeev:2015znc,Gursoy:2014aka,Voronyuk:2011jd,Deng:2012pc,Kharzeev:2007jp,Roy:2015coa,Alam:2021hje,Zhao:2019crj}.
Theoretically, these peak values of electromagnetic field are known to increase with the collision energy $\sqrt{s_{\text{NN}}}$\cite{Deng:2012pc,Panda:2024ccj}. Additionally, as the impact parameter increases, the event-averaged absolute values of the electromagnetic fields generally decrease, except for the $y$-component of the magnetic field, which initially rises before falling $\sim$ $b = 10-12$ fm~\cite{Tuchin:2013ie,Voronyuk:2011jd,Bzdak:2011yy,Deng:2012pc,Panda:2024ccj} at the initial time and also at the centre of the collision system. The vacuum solutions of the Maxwell's equation decays quite fast, but in the presence of a conducting medium these fields sustain longer~\cite{Kharzeev:2007jp,PhysRevC.107.034901}. It was also shown that in expanding charged fluids the strength of the fields are maintained for a more extended period as compared to the vacuum evolution~\cite{Dash:2023kvr}. However, at lower collision energies, electromagnetic fields, even without conductivity, are observed to decay more slowly compared to higher $\sqrt{s_{\text{NN}}}$ values. On the other hand, numerous theoretical advancements have been made to study the evolution of dissipative stresses in the QGP medium, including the effects of electromagnetic fields~\cite{STAR:2023jdd,Tuchin:2014hza,Das:2017qfi,Denicol:2018rbw,Denicol:2019iyh,Dubla:2020bdz,Das:2022lqh,Panda:2023akn,Inghirami:2019mkc,Panda:2020zhr,Panda:2021pvq,Ambrus:2022vif,Dash:2022xkz,Kushwah:2024zgd,Singh:2024leo,Singh:2023pwf,Palni:2024wdy}.
In most scenarios, only the contributions from spectator nucleons (nucleons that do not participate in the collision) are considered.
%All these features are primarily derived from the solutions of Maxwell's equations without considering the inputs from the fluid except~\cite{Dash:2023kvr}.
However, as the fields and fluids evolve concurrently, they exert mutual influence on each other. Consequently, ongoing efforts are focused on developing numerical magnetohydrodynamics (MHD) codes to comprehensively investigate the interactions between fluids and fields~\cite{Mayer:2024kkv,Mayer:2024dze,Nakamura:2022wqr,Nakamura:2022idq,Nakamura:2022ssn,Dash:2022xkz,Huang:2022qdn}.
 These studies offer unique opportunities to investigate the effects of electromagnetic fields at both the initial and final hadronic states~\cite{Satapathy:2021cjp,Dash:2020vxk,Singh:2023pwf,Ghosh:2022vjp}.

Although electromagnetic fields have been established theoretically to affect the QGP medium formed in heavy-ion collision systems, disentangling the effects of electromagnetic signals from experimental data using suitable observables remains a crucial research area for both theorists and experimentalists in the heavy-ion collision community. Several experimental and theoretical efforts have been made to probe these effects~\cite{Sun:2023rhh,Sun:2023adv,STAR:2023jdd,Alam:2021hje,ALICE:2019sgg}. 
Given the potential to confirm the influence of electromagnetic fields at lower energies, the Beam Energy Scan (BES) program at RHIC and the upcoming Facility for Antiproton and Ion Research (FAIR) at GSI could provide valuable insights. These experiments could also present an opportunity to study novel phenomena such as the chiral magnetic effect (CME)~\cite{Kharzeev:2004ey,Fukushima:2008xe,Kharzeev:1998kz,Huang:2015oca,ALICE:2023weh,STAR:2021mii,STAR:2022ahj,Hattori:2022hyo}, which has attracted significant theoretical and experimental attention.
Related phenomena such as chiral magnetic waves (CMW)\cite{STAR:2022zpv,Kharzeev:2010gd} and the chiral separation effect (CSE)\cite{Huang:2013iia} could also be observed in lower-energy heavy-ion collisions. As an experimental signature, charge-dependent two- and three-particle correlations in isobaric collisions (collisions involving nuclei with the same mass number but different atomic numbers) have been proposed to indicate the presence of the CME. However, separating the CME signal from potential background effects requires a comparison of experimental data with theoretical models that accurately account for the spatiotemporal evolution of electromagnetic fields. 
Other potential signals include the charge splitting of directed and elliptic flow of both lighter and heavier particles as a function of transverse momentum ($p_T$) and rapidity ($y$) in both large and small systems~\cite{He:2023eqx,Parida:2023ldu,Parida:2022lmt,STAR:2023wjl,Sun:2021psy,Oliva:2020mfr,Sun:2023adv,Das:2016cwd}, at RHIC and LHC energies.

  Since the study of electromagnetic fields is more suitable at lower energies, here we primarily focus on collisions at lower center-of-mass energies and examine the charge splitting of the elliptic flow, $\Delta v_2$, as a function of transverse momentum, $p_T$. In this study, we calculate the electromagnetic fields by considering the effects of both electric and chiral conductivity in a realistic setting. We observe that the electric fields decay more slowly than the magnetic fields, which consequently contribute more significantly to the splitting of the elliptic flow. 
Additionally, we aim to estimate the maximum strength of the fields on the hypersurface that reasonably describes the data. 
For this analysis, we focus on pions, as they could provide a cleaner probe compared to protons and kaons. This is because protons and kaons may have additional contributions from other conserved charges, such as baryon number and strangeness, making pions a more straightforward choice.

This paper is organized as follows: In Sec.~\eqref{sec:formulations}, we sequentially outline the steps taken to evaluate the dependence of electromagnetic fields on \( \Delta v_2 \) with respect to \( p_T \). We begin in Sec.~\eqref{blastwave} by introducing the model for the fluid velocity profile. Then, in Sec.~\eqref{cooperfreyformalism}, we detail the formula for calculating the elliptic flow using the Cooper-Frye prescription. In Sec.~\eqref{emfields}, we expand on the expressions for the electromagnetic fields used in this study and explain the sampling process employed to obtain realistic estimates. Following this, Sec.~\eqref{effectsofem} revisits the procedure for evaluating the drift velocity, incorporating the effects of electromagnetic fields. The results of our analysis are presented in Sec.~\eqref{results}, and finally, we conclude our findings in Sec.~\eqref{conclusion}. Throughout the paper, we adopt natural units, where \( \hbar = c = k_B = \epsilon_0 = \mu_0 = 1 \).

  \section{Formulation}
\label{sec:formulations}

In this section, we outline the sequential methodology adopted in our current study.

\begin{itemize}
    \item Since we mainly focus on lower to intermediate center-of-mass energy collisions, we begin by modeling the fluid velocity and temperature profile by calculating the elliptic flow \( v_2 \) at these energies, specifically focusing on Au+Au collisions with \( \sqrt{s_{NN}} = 27\,\text{GeV} \) within the 0-80\% centrality range and comparing the results with those reported by the STAR collaboration~\cite{STAR:2013ayu} to extract the unknown parameters. 
Here, we employ the blastwave method to parameterize the fluid four-velocity, based on the assumption that the fluid flow profile is characterized by boost invariance and azimuthal symmetry. The calculation is performed using the Cooper-Frye prescription, where we utilize the equilibrium distribution function for pions (the particles of interest here). In this process, crucial parameters such as \( u_0 \) (initial velocity), \( c_2 \) (eccentricity parameter), and \( T_f \) (freeze-out temperature) are extracted. Additionally, we determine the hypersurface radius \( R \).

    \item With the fluid profile set, we proceed to calculate the electromagnetic field components at any given lattice point defined by the coordinates \( (t, x, y, z) \). This calculation utilizes the expressions provided in Eqs.~\eqref{eq:eq8} and \eqref{eq:eq9}, which are based on the assumption that the chiral conductivity \( \sigma_{\chi} \) is much smaller than the electrical conductivity \( \sigma \) (\( \sigma_{\chi} \ll \sigma \))~\cite{Li:2016tel, Siddique:2021smf, Siddique:2022ozg}. In Sec.~\eqref{emfields}, we detail the formulas and the setup of the colliding nuclei to calculate all the electromagnetic field components.

    \item Armed with the flow profile of the fluid and the electromagnetic fields, we proceed to investigate the impact of these fields by computing the drift velocity associated with the electromagnetic fields generated in such colliding systems. This drift velocity is then added perturbatively to the fluid velocity using the relativistic velocity addition rule, under the assumption that the drift velocity \( \vec{v} \) is significantly smaller than the fluid velocity \( \vec{u} \) (\( \vec{v} \ll \vec{u} \)), which is verified across the entire hypersurface.
The specifics of this perturbative approach are revisited in Sec.~\eqref{effectsofem}, drawing on methodologies from~\cite{Gursoy:2014aka}. This addition allows us to probe the subtle influences of electromagnetic fields on the fluid dynamics.

    \item Subsequently, we employ the Cooper-Frye prescription once again to calculate the elliptic flow for both particles and antiparticles, taking into account the effects of both the fluid and the fields. This computation enables us to determine \( \Delta v_2 \), defined as the difference between the elliptic flow of negatively charged pions (\( v_2^{\pi^{-}} \)) and positively charged pions (\( v_2^{\pi^{+}} \)) as a function of transverse momentum \( p_T \). This step is crucial for analyzing the charge-dependent flow patterns that arise due to the presence of electromagnetic fields.

    \item Finally, we compare our theoretical predictions with experimental data to validate our results. This comparison allows us to estimate the maximum field strengths that best match the experimental observations. Additionally, we discuss the effects of individual field components and highlight potential avenues for future research and investigation.

\end{itemize}

\section{Fluid Velocity modelling}\label{blastwave}
In high-energy heavy-ion collisions, a widely used parametrization of the fluid four-velocity is inspired by the Bjorken model, which describes boost-invariant expansion along the longitudinal direction. However, in this study, we focus on a radial velocity profile, assuming boost invariance, and allowing for expansion in the transverse plane.

We utilize the Milne coordinates \(\left(\tau, \eta, r, \phi\right)\) with the metric \(g_{\mu\nu} = \mathrm{diag}\left(1, -\tau^2, -1, -r^2\right)\)~\cite{Teaney:2003kp,Panda:2023akn}.
Here $\tau$ is the proper time and $r$ is the radial distance on the transverse plane.
The transformation between Cartesian and Milne coordinates is defined as:

\begin{eqnarray}
\tau &=& \sqrt{t^2 - z^2}, \nonumber \\
\eta &=& \tanh^{-1}{\left(\frac{z}{t}\right)}, \nonumber \\
r &=& \sqrt{x^2 + y^2}, \nonumber \\
\phi &=& \arctan2(y, x).
\end{eqnarray}

The parameterized form of the velocity four-vector, incorporating longitudinal boost-invariance, is given by:

\begin{eqnarray}
u^{r} &=& u_0 \frac{r}{R_f} \left[1 + 2\sum_{n=1}^{\infty} c_n \cos{\left(n\left[\phi - \psi_n\right]\right)}\right] \Theta\left(R_f - r\right), \nonumber \\
u^{\phi} &=& u^{\eta} = 0, \nonumber \\
u^{\tau} &=& \sqrt{1 + \left(u^r\right)^2}.
\label{parametrisedvelocity}
\end{eqnarray}

Here, \(u^{\tau}, u^{r}, u^{\phi},\) and \(u^{\eta}\) are the components of the fluid four-velocity. The parameters \(u_0\) and \(c_n\) are free parameters used to reproduce the transverse momentum spectra and the flow harmonics of charged hadrons. The Heaviside function \(\Theta\left(R_f - r\right)\) ensures that \(u^{r} = 0\) when \(r > R_f\), where \(R_f\) represents the radius of the freeze-out hypersurface. The azimuthal angle in coordinate space is denoted by \(\phi\), and \(\psi_n\) is the \(n\)-th order participant plane angle. Since event-by-event fluctuations are not considered in this work, we set \(\psi_n = 0\), aligning the minor axis of the participant planes with the direction of the impact parameter.
The temperature on the freeze-out hypersurface is parameterized as:

\begin{eqnarray}
T\left(\tau, \eta, r, \phi\right) &=& T_0 \Theta\left(R_f - r\right),
\end{eqnarray}

where \(T_0\) denotes the temperature at the freeze-out hypersurface.

\subsection{Cooper-frey formalism} \label{cooperfreyformalism}
As mentioned earlier, the invariant yield of hadrons is obtained using the Cooper-Frye formula~\cite{Cooper:1974mv}, which assumes that the freeze-out hypersurface is a timelike vector \( d\Sigma_{\mu} \), given by \( \left(\tau d\eta dr rd\phi,0,0,0\right) \). The invariant yield is expressed as:

\begin{eqnarray} \label{eq:invYield}
\frac{dN}{d^2p_{T} dy} &=& \frac{\mathcal{G}}{\left(2\pi\right)^3} \int p^{\mu} d\Sigma_{\mu} f(x,p),
\end{eqnarray}

where \( f(x,p) \) is the single-particle distribution function, and \( x \) and \( p \) are the position and momentum four-vectors of the particles, respectively. \( \mathcal{G} \) represents the degeneracy factor.

In rapidly expanding fireballs, the system may not achieve local thermal equilibrium. As a result, the single-particle distribution function needs to incorporate deviations from equilibrium when calculating hadron yields using the Cooper-Frye prescription. These deviations account for factors such as viscosity and electromagnetic fields. However, in this study, we concentrate exclusively on the effects of electromagnetic fields, disregarding the influence of viscosity.

Furthermore, the effects of electromagnetic fields can be incorporated in two broad ways: either by including an off-equilibrium correction term or by adding a drift velocity correction to the existing fluid velocity. In this study, we adopt the latter approach, as it is more suitable when both spectators and participants significantly influence the medium, making it preferable to directly modify the fluid velocity profile.

Further, the \( n \)-th order flow coefficient is calculated using the standard formula that is a function of transverse momentum ($p_T$) and momentum rapidity (y):

\[
v_n(p_T, y) = \frac{\int_0^{2\pi} d\phi \cos(n\phi) \frac{dN}{d^2p_T \, dy}}{\int_0^{2\pi} d\phi \frac{dN}{d^2p_T \, dy}}.
\]

To determine all the unknown parameters, the above formula is applied using the equilibrium distribution function \( f_0 = \frac{1}{e^{\beta (u \cdot p) - \alpha} - 1} \), where \( \beta = \frac{1}{T} \) is the inverse temperature, \( T \) is the temperature, \( u^{\mu} \) is the fluid four-velocity, and \( \alpha = \frac{\mu}{T} \) represents the ratio of the chemical potential \( \mu \) to the temperature (T). This formulation coupled with the previously defined timelike hypersurface for evaluating the spectra, is used to evaluate the elliptic flow for pions. 
\begin{figure}[H]
	    \centering
	    \includegraphics[width=1\linewidth]{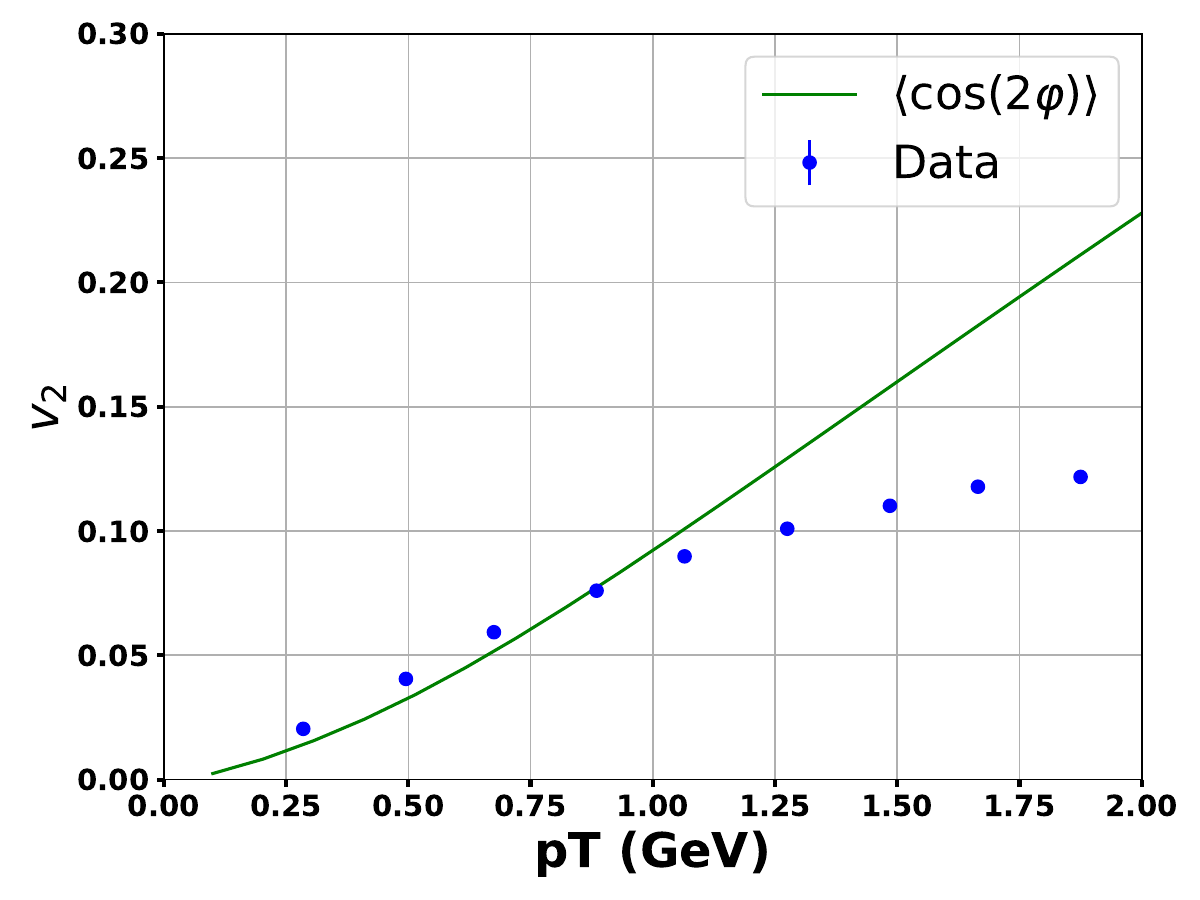}
	    \caption{(Color Online) Fit of the elliptic flow $v_2$ as a function of $p_T$, using data from the Au + Au collision at STAR experiment~\cite{STAR:2013ayu}, with $\sqrt{s_{\text{NN}}} = 27$ GeV at mid-rapidity ($y=0$), and 0-80\% centrality.
}
	    \label{v2}
	\end{figure}

The parameters: freeze-out temperature \( T = 0.0785 \) GeV, radius \( R_f = 9.0 \) fm, maximum fluid four-velocity \( u_0 = 0.65 \), proper time \( \tau = 4 \) fm, and spatial eccentricity \( c_2 = 0.04 \), provide a good fit to the experimental data for the pion elliptic flow \( v_2 \) within the 0-80\% centrality range at \( \sqrt{s_{\text{NN}}} = 27 \) GeV. The fit for \( v_2 \) is shown in Fig.~\eqref{v2}.

The selection of this particular energy is entirely arbitrary and unbiased. Nevertheless, any energy within the range \( \sqrt{s_{\text{NN}}} = 62.4 \, \text{GeV} \) to \( 7.7 \, \text{GeV} \) could have been investigated. However, in the latter part of this paper, we will also extrapolate our findings to cover these energy ranges.

In the QCD phase diagram, exploring lower energies helps probe the region of higher, non-zero chemical potential, indicating that the chemical potential is energy-dependent and becomes significant at lower center-of-mass energies. The presence of a non-zero chemical potential can create an imbalance between particles and antiparticles, potentially causing a splitting in their flow. However, in this study, we focus solely on the effects of electromagnetic fields and, therefore, set the chemical potential to zero. 

\section{Electromagnetic field configurations
}\label{emfields}
Before setting up the calculations to determine the drift velocity due to the presence of electromagnetic fields, we first review the electromagnetic field configurations employed in the current analysis. We begin by presenting the formula for all components of the fields, assuming finite and constant electrical and chiral conductivities, which are used in this study. Several previous works~\cite{Gursoy:2014aka,Siddique:2021smf,Siddique:2022ozg} have extensively studied these fields, that are evaluated by considering that all source charges propagate along the \( z \)-axis and by employing the Green's function method in cylindrical coordinates~\cite{Li:2016tel,Siddique:2021smf,Siddique:2022ozg}. The expressions for the electromagnetic fields are as follows:

\begin{align}
B_{\phi}(t,\textbf{x}) & =\frac{Q}{4\pi}\frac{v\gamma x_{\mathrm{T}}}{\Delta^{3/2}}\left(1+\frac{\sigma v\gamma}{2}\sqrt{\Delta}\right)e^{A},\nonumber \\
B_{r}(t,\textbf{x}) & =-\sigma_{\chi}\frac{Q}{8\pi}\frac{v\gamma^{2}x_{\mathrm{T}}}{\Delta^{3/2}}e^{A}\left[\gamma\left(vt-z\right)+A\sqrt{\Delta}\right],\nonumber \\
B_{z}(t,\textbf{x}) & =\sigma_{\chi}\frac{Q}{8\pi}\frac{v\gamma}{\Delta^{3/2}}e^{A}\Big[\Delta\left(1-\frac{\sigma v\gamma}{2}\sqrt{\Delta}\right)\nonumber \\
 & \;\;\;\;+\gamma^{2}\left(vt-z\right)^{2}\left(1+\frac{\sigma v\gamma}{2}\sqrt{\Delta}\right)\Big],\label{eq:eq8}
\end{align}
in which $\Delta$ and $A$ are defined as $\Delta\equiv\gamma^{2}\left(vt-z\right)^{2}+x_{\mathrm{T}}^{2}$
and $A\equiv\left(\sigma v\gamma/2\right)\left[\gamma\left(vt-z\right)-\sqrt{\Delta}\right]$, with $x_{\mathrm{T}}$ being the magnitude of the transverse coordinate $x_{\mathrm{T}}=\sqrt{x^{2}+y^{2}}$; and 
\begin{align}
E_{\phi}(t,\textbf{x}) & = \sigma_{\chi} \frac{Q}{8\pi} \frac{v^{2} \gamma^{2} x_{\mathrm{T}}}{\Delta^{3/2}} e^{A} \left[\gamma(vt-z) + A \sqrt{\Delta}\right], \nonumber \\
E_{r}(t,\textbf{x}) & = \frac{Q}{4\pi} e^{A} \Bigg\{ \frac{\gamma x_{\mathrm{T}}}{\Delta^{3/2}} \left(1 + \frac{\sigma v \gamma}{2} \sqrt{\Delta}\right) \nonumber \\
 & \;\;\;\; - \frac{\sigma}{v x_{\mathrm{T}}} e^{-\sigma(t-z/v)} \left[1 + \frac{\gamma(vt-z)}{\sqrt{\Delta}}\right] \Bigg\}, \nonumber
\end{align}
\begin{align}
E_{z}(t,\textbf{x}) & = \frac{Q}{4\pi} \left\{ - \frac{e^{A}}{\Delta^{3/2}} \left[\gamma(vt-z) + A \sqrt{\Delta} + \frac{\sigma \gamma}{v} \Delta \right] \right. \nonumber \\
 & \;\;\;\;\left. + \frac{\sigma^{2}}{v^{2}} e^{-\sigma(t-z/v)} \Gamma\left(0, -A\right) \right\}, \label{eq:eq9}
\end{align}

with $\Gamma\left(0,-A\right)$ being the incomplete gamma function
defined as $\Gamma\left(a,z\right)=\int_{z}^{\infty}t^{a-1}\exp\left(-t\right)\,dt$.
$Q$ in the above expressions is the charge, $\sigma$ and $\sigma_{\chi}$ are electrical and chiral conductivities respectively, with the constrain that $\sigma_{\chi}$ $\ll$ $\sigma$.
The expressions here that are in cylindrical coordinates are transformed to the cartesian co-ordinates using suitable trasformations.

The system under consideration consists of two heavy nuclei (Gold, Au), whose centers are positioned at (\( \pm \frac{b}{2}, 0, 0 \)). The nucleus centered at \( x_0 = -\frac{b}{2} \) moves along the negative \( z \)-direction, while the other nucleus moves along the positive \( z \)-direction. 
Before calculating the electromagnetic fields at any desired point on the space-time grid, the positions of the protons in the nuclei must first be sampled. This serves as the input for the field calculations. The sampling is performed using the Woods-Saxon distribution, which is defined as:

\begin{equation} \label{woodsaxon}
    \rho(r) = \frac{\rho_0}{1 + e^{(r - R)/a}},
\end{equation}

where \( \rho_0 \) is the density at the center of the nucleus, taken to be 0.16 \( \text{fm}^{-3} \) for Au, \( R \) is the radius (6.34 fm for Au), and \( a \) is the diffuseness parameter (0.54 fm)~\cite{Miller:2007ri}.

\begin{figure}[H]
    \centering
    \includegraphics[width=1\linewidth]{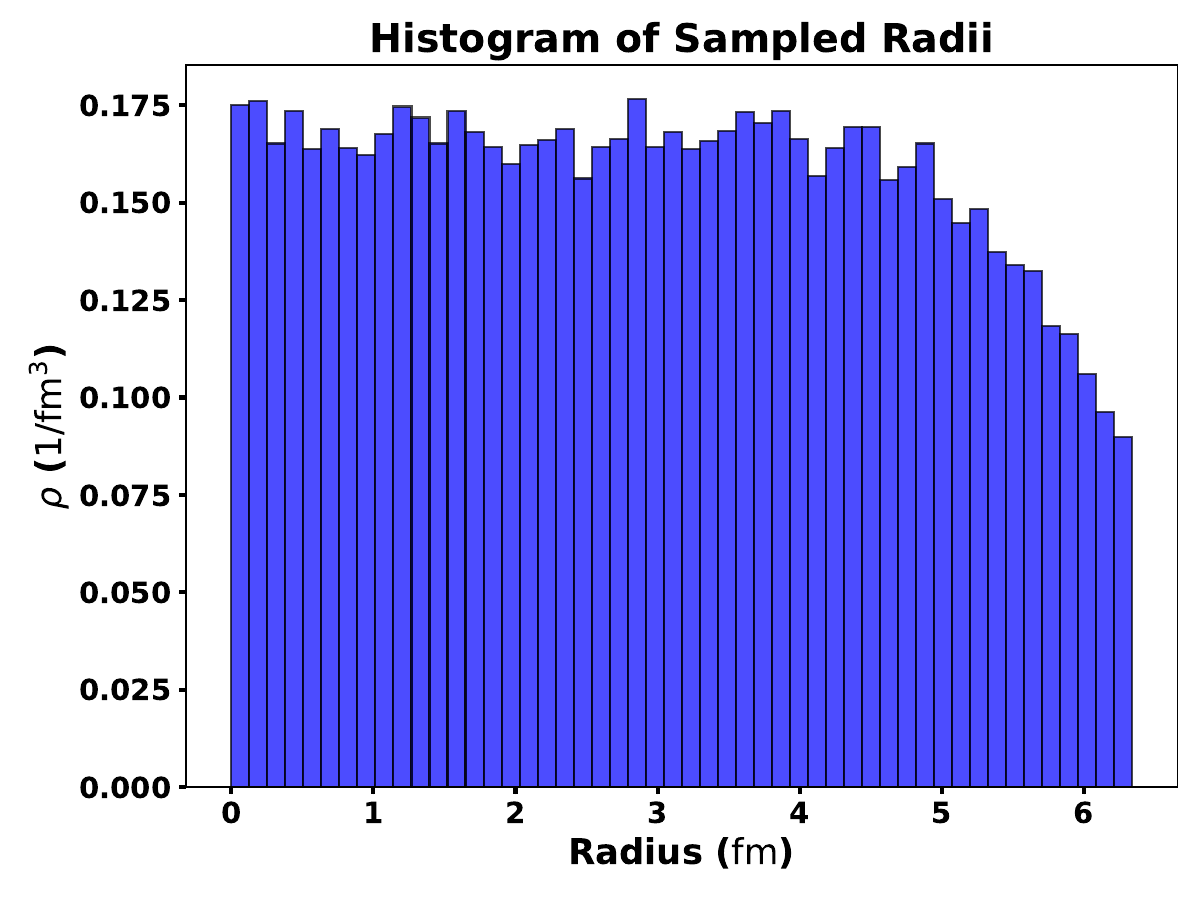}
    \caption{(Color Online) Sampled radii from 1000 events (158 in each event) that were used as positions for protons in the nucleus in 1000 different events, thus smoothening the proton distribution according to the Wood-Saxon profile on an average.}
    \label{woodsaxon1}
\end{figure}
Protons are considered fixed on the transverse plane, with motion along the \( z \)-direction. Both nuclei move along the z direction with \( \pm vt \), where \( v \) is the velocity along the beam direction, determined from the center-of-mass energy:

\[
v = \sqrt{1 - \frac{4 m^2}{(\sqrt{s_{\text{NN}}})^2}}.
\]

Monte Carlo methods are used to sample transverse plane coordinates according to Eq.~\eqref{woodsaxon}. In each event, 158 protons (79 per nucleus) are sampled, with centers at \( (\pm \frac{b}{2}, 0, 0) \). Once the positions are sampled, the inter-nucleon transverse distance \( r_{\bot} = \sqrt{(x_p - x_T)^2 + (y_p - y_T)^2} \) is calculated, where $(x_T, y_T)$ and $(x_p, y_p)$ are the coordinates of the target and projectiles respectively. If \( r_{\bot} \leq r_c \), where

\[
r_{c} = \sqrt{\frac{\sigma_{NN}}{\pi}},
\]

protons are considered participants; otherwise, they are spectators. The $\sigma_{NN}$ value is taken from~\cite{Panda:2024ccj} by extracting the value at a particular $\sqrt{s_{\text{NN}}}$ after extrapolating the fit result with the experimentally measured values. Participants once identified are assigned their sampled positions and have non-zero electrical and chiral conductivity (for the current study, we use \( \sigma = 0.023 \, \text{fm}^{-1} \) and \( \sigma_{\chi} = 0.004 \, \text{fm}^{-1} \)), while spectators have zero conductivity. This process is repeated for 1000 events to achieve smooth electromagnetic field configurations on the transverse plane or on the hypersurface of interest. The sampled position distribution for 1000 events is shown in Fig.~\eqref{woodsaxon1}. Although baryon stopping may influence the results at these energies~\cite{Panda:2024ccj,Taya:2024wrm}, it is neglected for now.

\begin{figure*}
    \centering
    % First row - Electric Fields with Conductivity
    \begin{subfigure}{0.35\linewidth}
        \centering
        \includegraphics[width=1\linewidth]{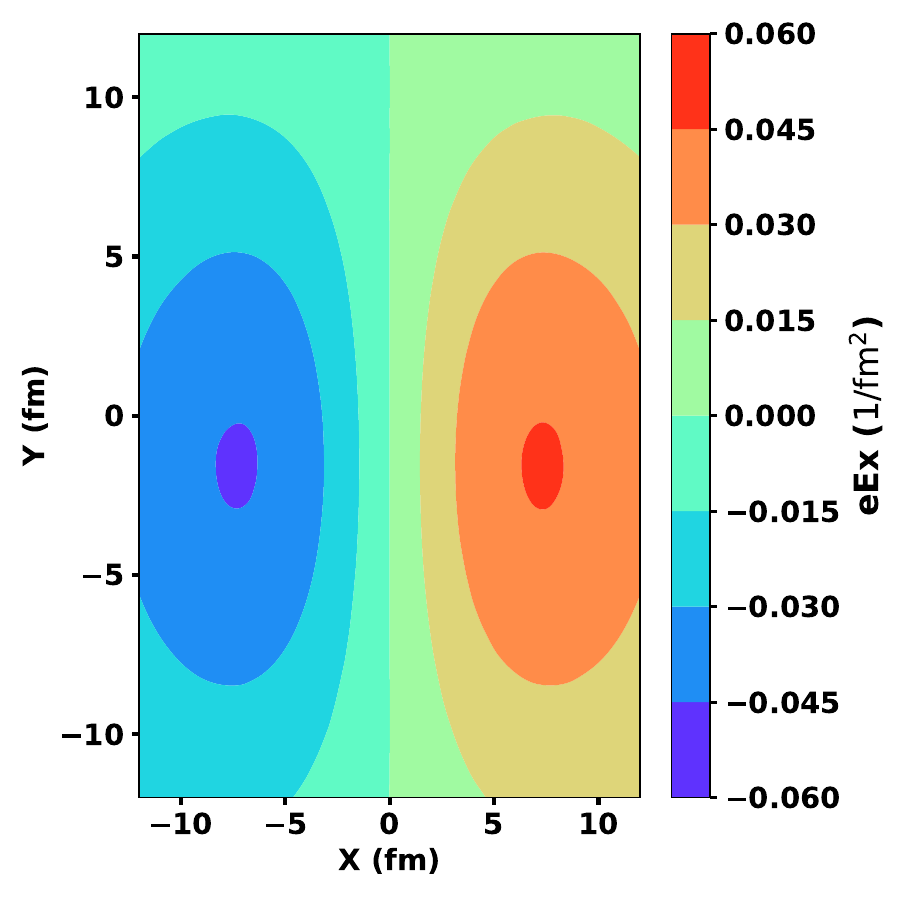}
        \label{fig:eyc}
    \end{subfigure}
    \hspace{0.02\linewidth}
    \begin{subfigure}{0.35\linewidth}
        \centering
        \includegraphics[width=1\linewidth]{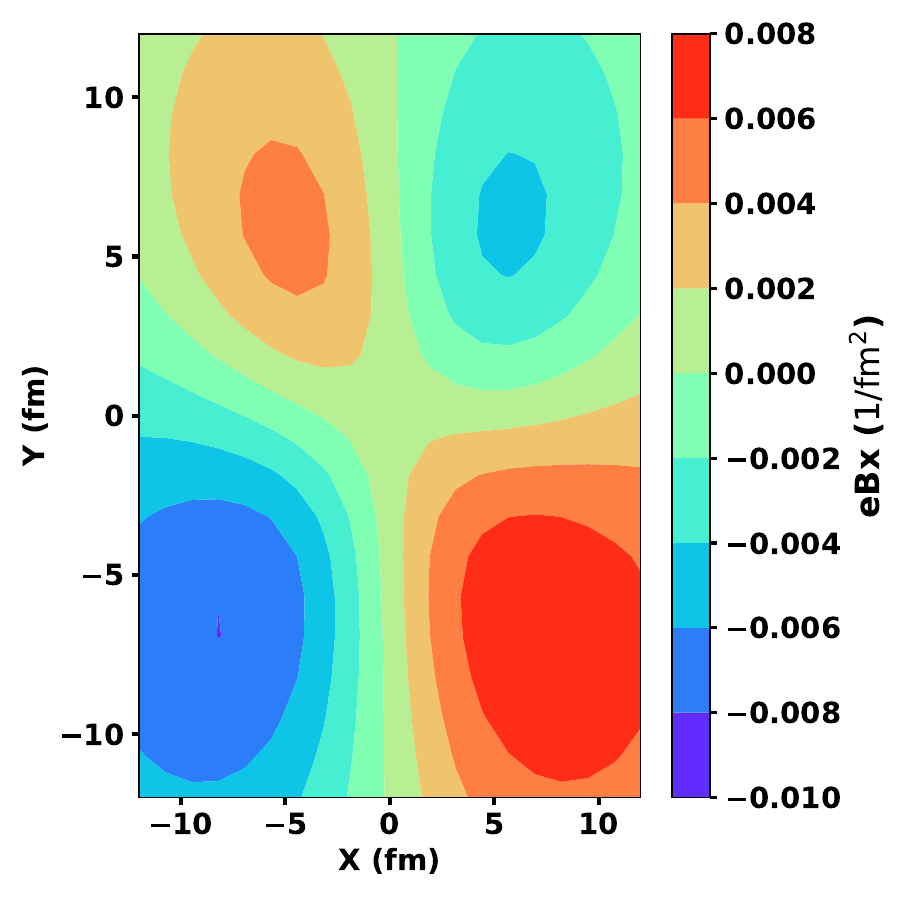}
        \label{fig:exc}
    \end{subfigure}

    % Second row - Magnetic Fields with Conductivity
    \begin{subfigure}{0.35\linewidth}
        \centering
        \includegraphics[width=1\linewidth]{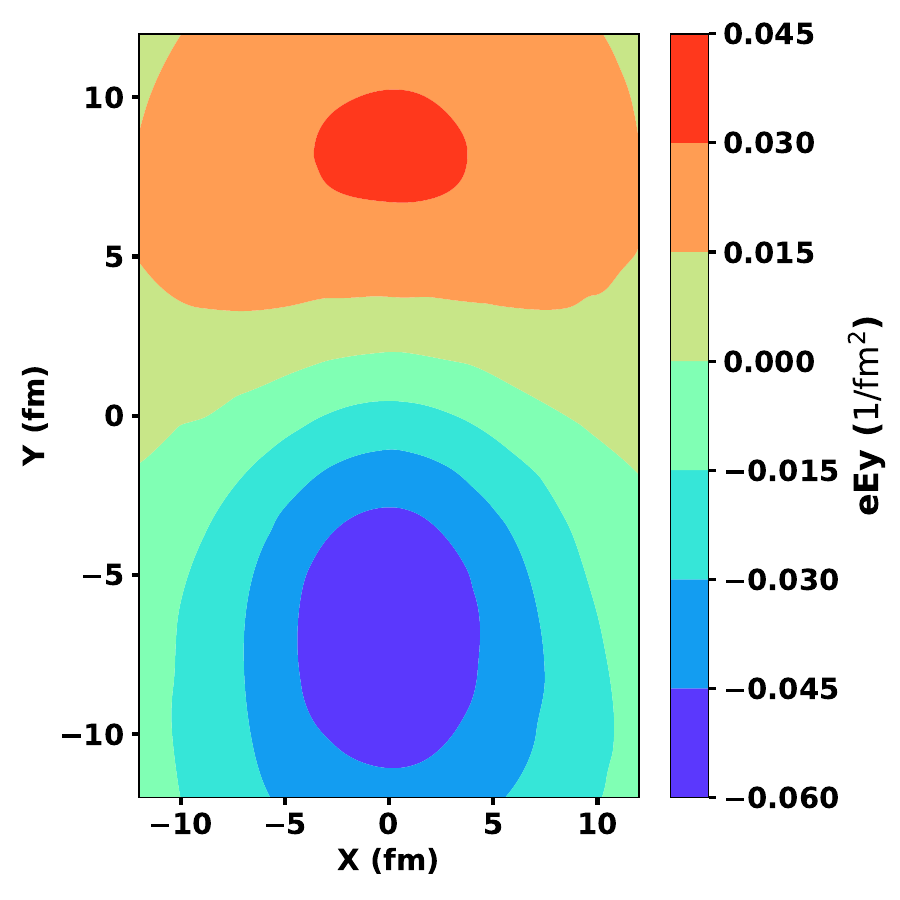}
        \label{fig:byc}
    \end{subfigure}
    \hspace{0.02\linewidth}
    \begin{subfigure}{0.35\linewidth}
        \centering
        \includegraphics[width=1\linewidth]{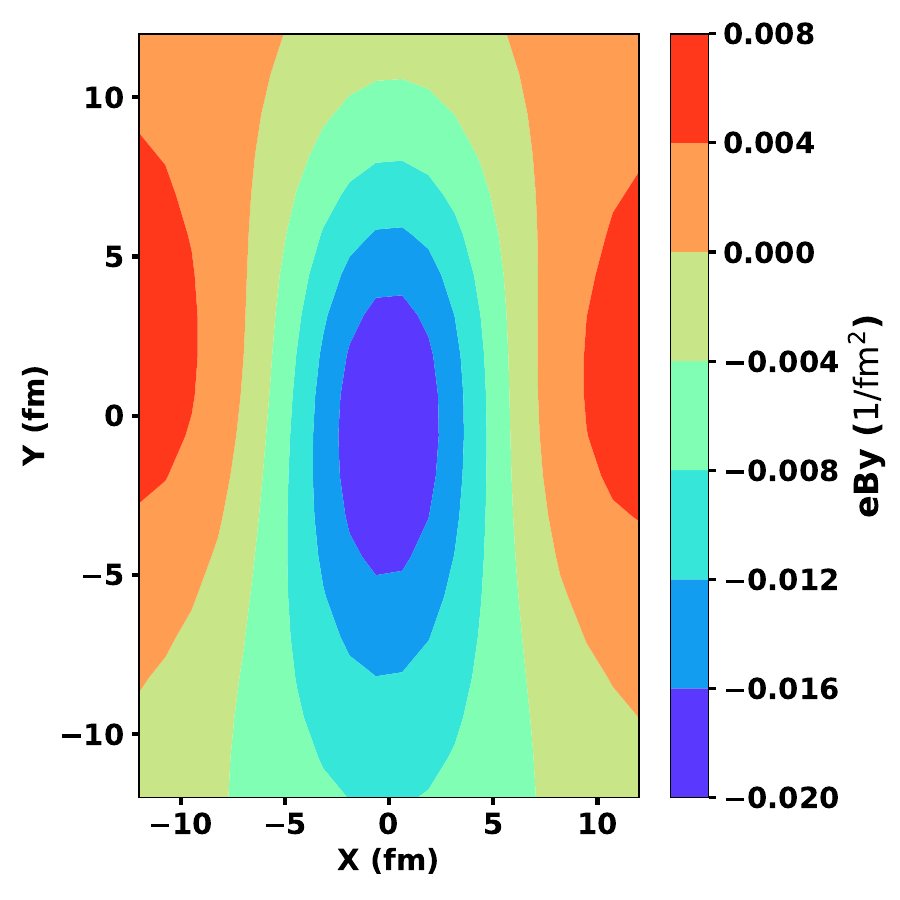}
        \label{fig:bxc}
    \end{subfigure}

    % Third row - E_z and B_z Fields
    \begin{subfigure}{0.35\linewidth}
        \centering
        \includegraphics[width=\linewidth]{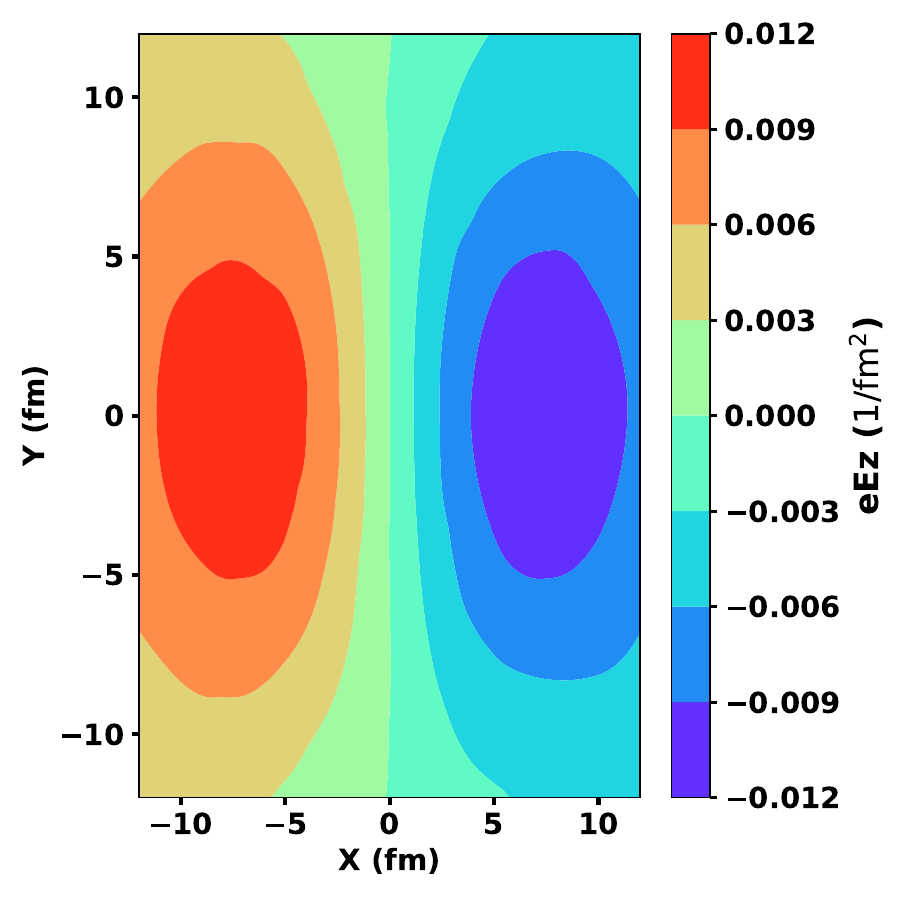}
        \label{fig:ez}
    \end{subfigure}
    \hspace{0.02\linewidth}
    \begin{subfigure}{0.35\linewidth}
        \centering
        \includegraphics[width=\linewidth]{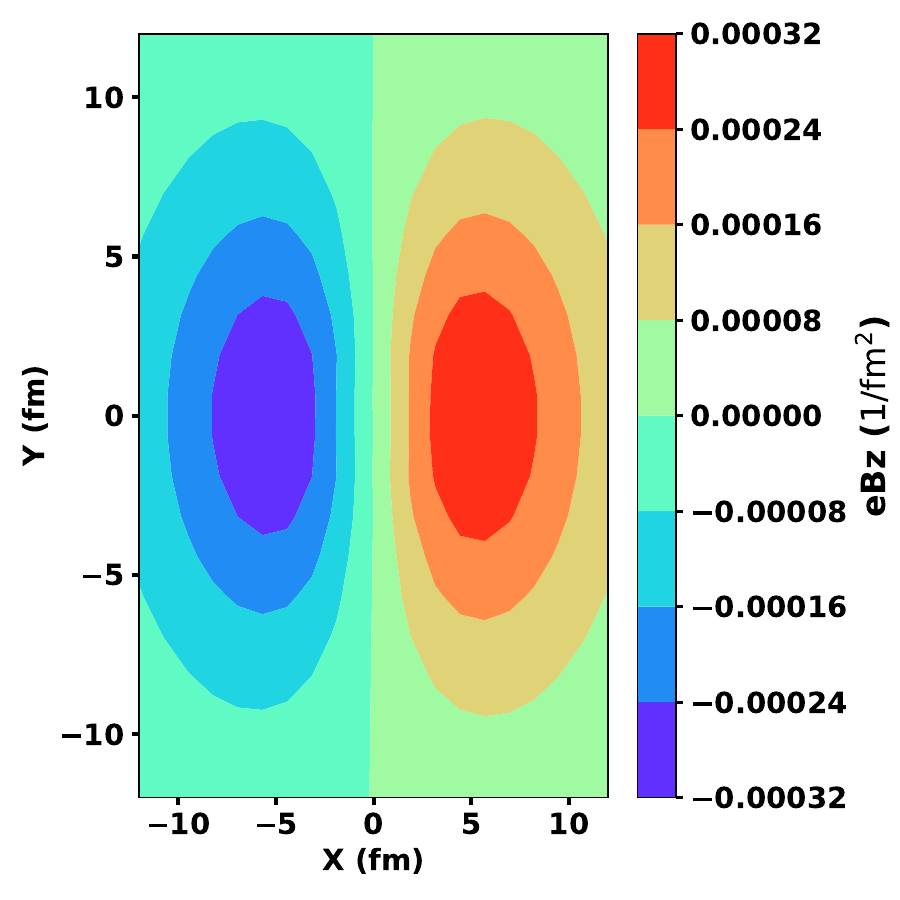}
        \label{fig:bz}
    \end{subfigure}

    \caption{(Color Online) Electric and magnetic fields with conductivity are plotted on the transverse plane as follows: 
(a) The top row shows \( eE_x \) and \( eB_x \), 
(b) the middle row shows \( eE_y \) and \( eB_y \), 
and (c) the last row shows \( eE_z \) and \( eB_z \), 
respectively, at \( (t,z) = (1,0) \) for Au+Au collisions at \( \sqrt{s_{\text{NN}}} = 27 \) GeV with an impact parameter of \( b = 3 \) fm along the \( x \)-direction. 
The calculations include finite conductivity with \( \sigma = 0.023 \) fm\(^{-1}\) and \( \sigma_{\chi} = 0.004 \) fm\(^{-1} \).
}
    \label{fig:all_fields}
\end{figure*}

With the above setup, we now calculate the electromagnetic fields using the formulas in Eq.~\eqref{eq:eq8} and~\eqref{eq:eq9}. Fig.~\eqref{fig:all_fields} shows the contour plots of the electromagnetic fields calculated on a transverse \( (X,Y) \) plane, with conductivity values \( \sigma = 0.023 \, \text{fm}^{-1} \) and \( \sigma_{\chi} = 0.004 \, \text{fm}^{-1} \) for \( \sqrt{s_{\text{NN}}} = 27 \) GeV, at \( (t,z) = (1,0) \) and an impact parameter \( b = 3 \) fm. In this setup, using the given target and projectile nuclei, we obtain the correct sign for the electromagnetic fields, as observed from the \( eB_y \) plot which yields a negative value at the center of the collision system. Here we can see that in the presence of conductivity, the symmetry of the fields is somewhat compromised, a phenomenon that is further discussed in details in~\cite{Li:2016tel,Siddique:2021smf,Siddique:2022ozg}.

\section{Effect of EM fields}\label{effectsofem}

With the electromagnetic fields and fluid profile established, we proceed to calculate the drift velocity in the rest frame of the fluid, as opposed to the center-of-mass frame where the electromagnetic fields are already known. The equation governing the drift velocity, assuming a stationary current condition, is given by:

\begin{eqnarray}\label{current}
    m \frac{d\mathbf{v}}{dt} = q\mathbf{v} \times \mathbf{B} + q\mathbf{E} - \mu m \mathbf{v} = 0,
\end{eqnarray}

where \(m\) is the mass of the charged fluid, on which the external forces (Lorentz forces) due to the first two terms on the left-hand side (\(q\mathbf{v} \times \mathbf{B} + q\mathbf{E}\)) act, while the last term represents the drag force with \(\mu\) being the drag coefficient and \(q\) the charge associated with the particle density. 
Following the approach in~\cite{Gursoy:2014aka}, we first transform the magnetic and electric fields, given in Eq.~\eqref{eq:eq8} and Eq.~\eqref{eq:eq9}, from the center-of-mass frame to the rest frame of the fluid using a Lorentz boost with velocity \(-\vec{u}\) at all points on the hypersurface. The drift velocity is then solved at each spacetime point, with \(q=1\) for particles and \(q=-1\) for anti-particles.

The quantity \(\mu m\) in Eq.~\eqref{current} is, in general, temperature dependent. However, we consider the same value as used in Ref.~\cite{Gursoy:2014aka} for simplicity.  Once the drift velocity is obtained for both particles and anti-particles, we perform a second Lorentz transformation given by \(\Lambda(\vec{u})\) on the drift velocities thus obtained using the suitable relativistic velocity addition formula to obtain the total velocity \(V^\mu_{\pm}\) in the center-of-mass frame, where \(V^\mu_+\) and \(V^\mu_-\) represent the velocities for positively and negatively charged particle densities, respectively.

At this stage, we verify that our assumption \(|\vec{v}|/|\vec{u}| \ll 1\) holds by calculating the Lorentz factor difference for the total velocity \(V^\mu_\pm\) and the fluid velocity \(u^\mu\). This difference is found to be small (of the order \(\sim 0.001\)) or less throughout the \((x, y, z)\) space, which is consistent with our assumptions.

With this, we are now ready to investigate the impact of the electromagnetic field on the elliptic flow \(v_2\). We employ the Cooper-Frye prescription discussed earlier, but now using the transformed velocities \(V^{\mu}_+\) for particles and \(V^{\mu}_-\) for anti-particles, instead of the original fluid velocity \(u^{\mu}\). Using these transformed velocities, we calculate the change in elliptic flow, \(\Delta v_2\), for pions at \(\sqrt{s_{\text{NN}}} = 27\) GeV, which is the central focus of this study.

\section{Results }\label{results}
\begin{figure}
    \centering
    \begin{subfigure}[b]{\linewidth}
        \centering
        \includegraphics[width=\linewidth]{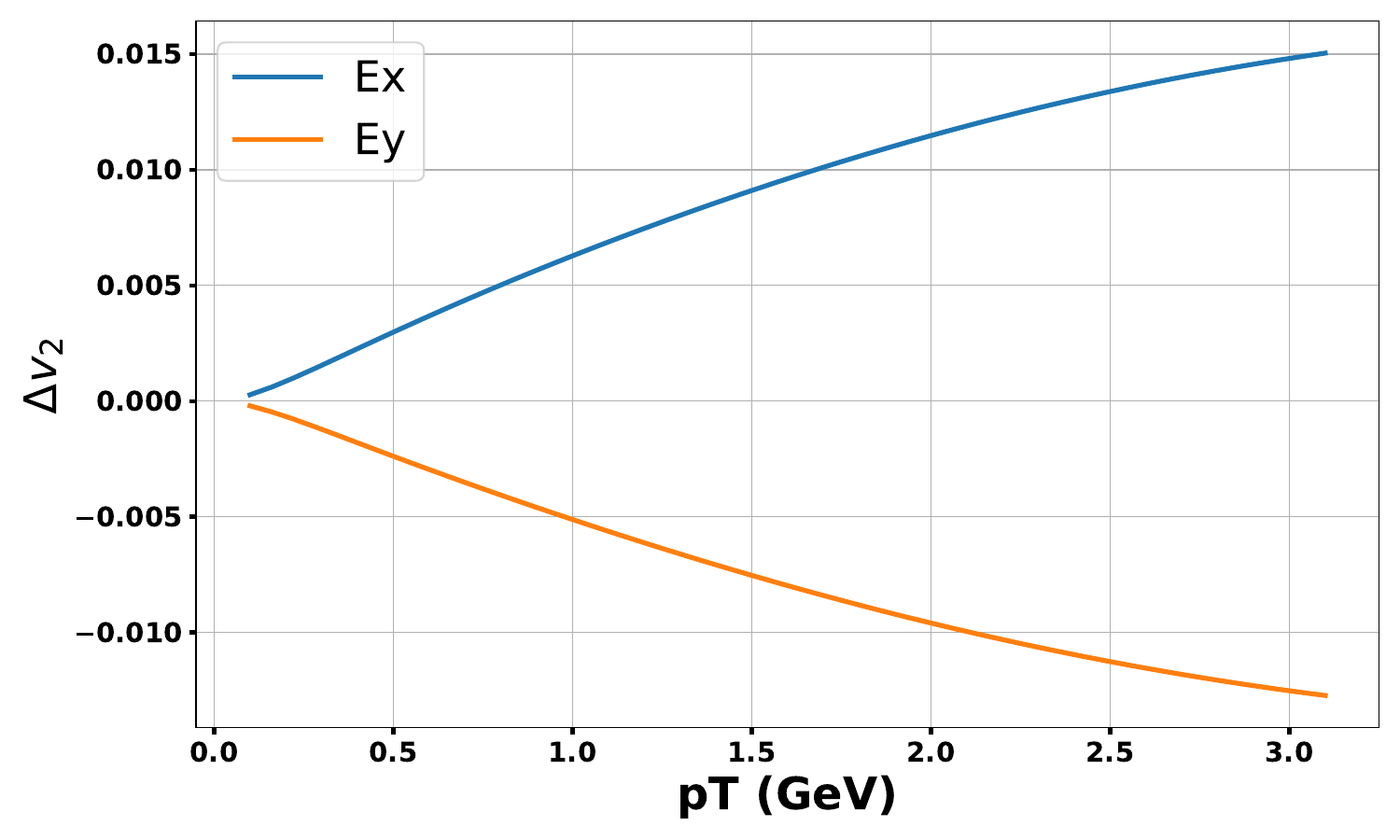}
        \caption{}
        \label{fig:rato1}
    \end{subfigure}
    \vskip 0.3cm
    \begin{subfigure}[b]{\linewidth}
        \centering
        \raisebox{0.3\baselineskip}[0pt][0pt]{\hspace*{0.8\linewidth}\scalebox{1.0}{$1 \times 10^{-5}$}}
        \includegraphics[width=\linewidth, trim=0 0 0 20, clip]{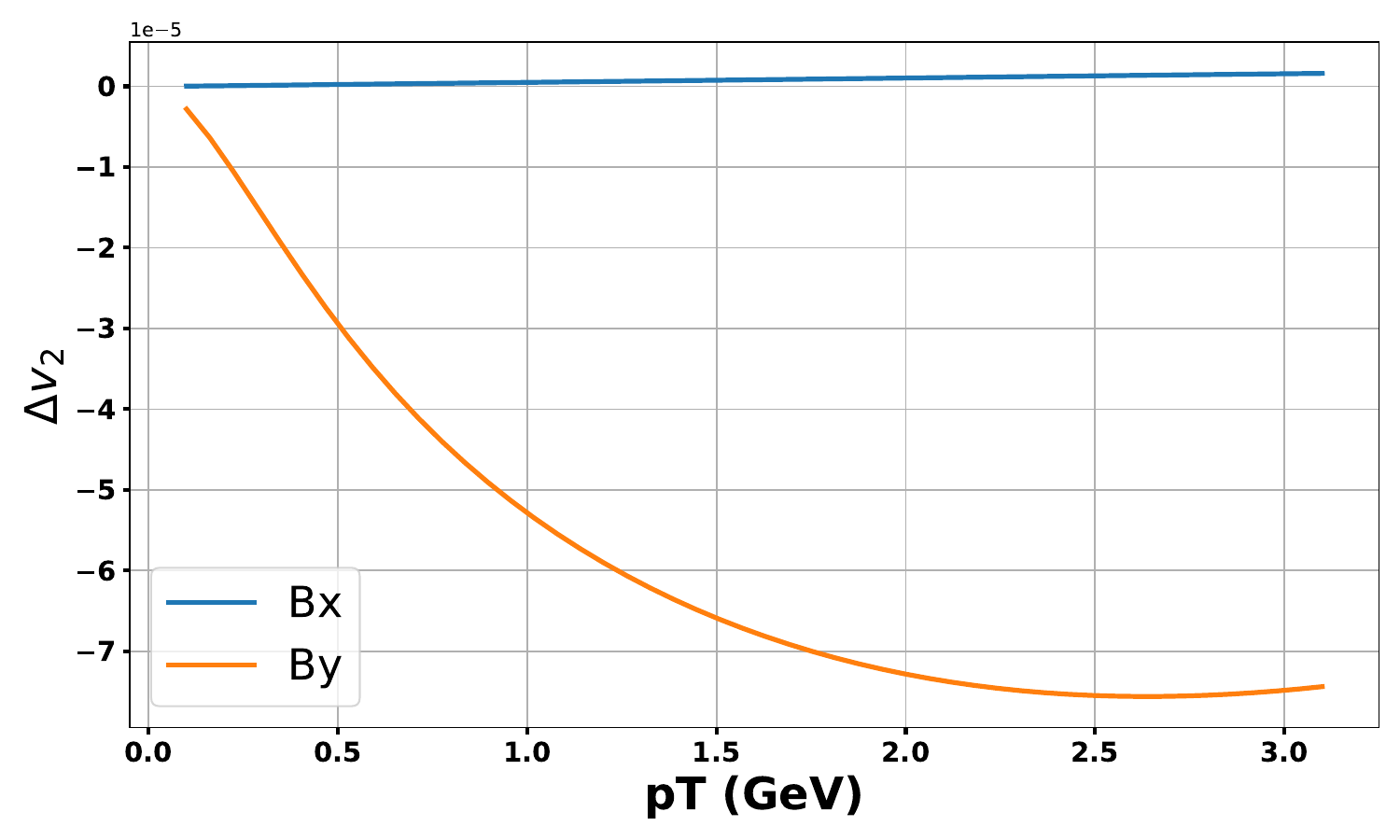}
        \caption{}
        \label{fig:rato2}
    \end{subfigure}
    \vskip 0.3cm
    \begin{subfigure}[b]{\linewidth}
        \centering
        \raisebox{0.3\baselineskip}[0pt][0pt]{\hspace*{0.8\linewidth}\scalebox{1.0}{$1 \times 10^{-5}$}}
        \includegraphics[width=\linewidth, trim=0 0 0 20, clip]{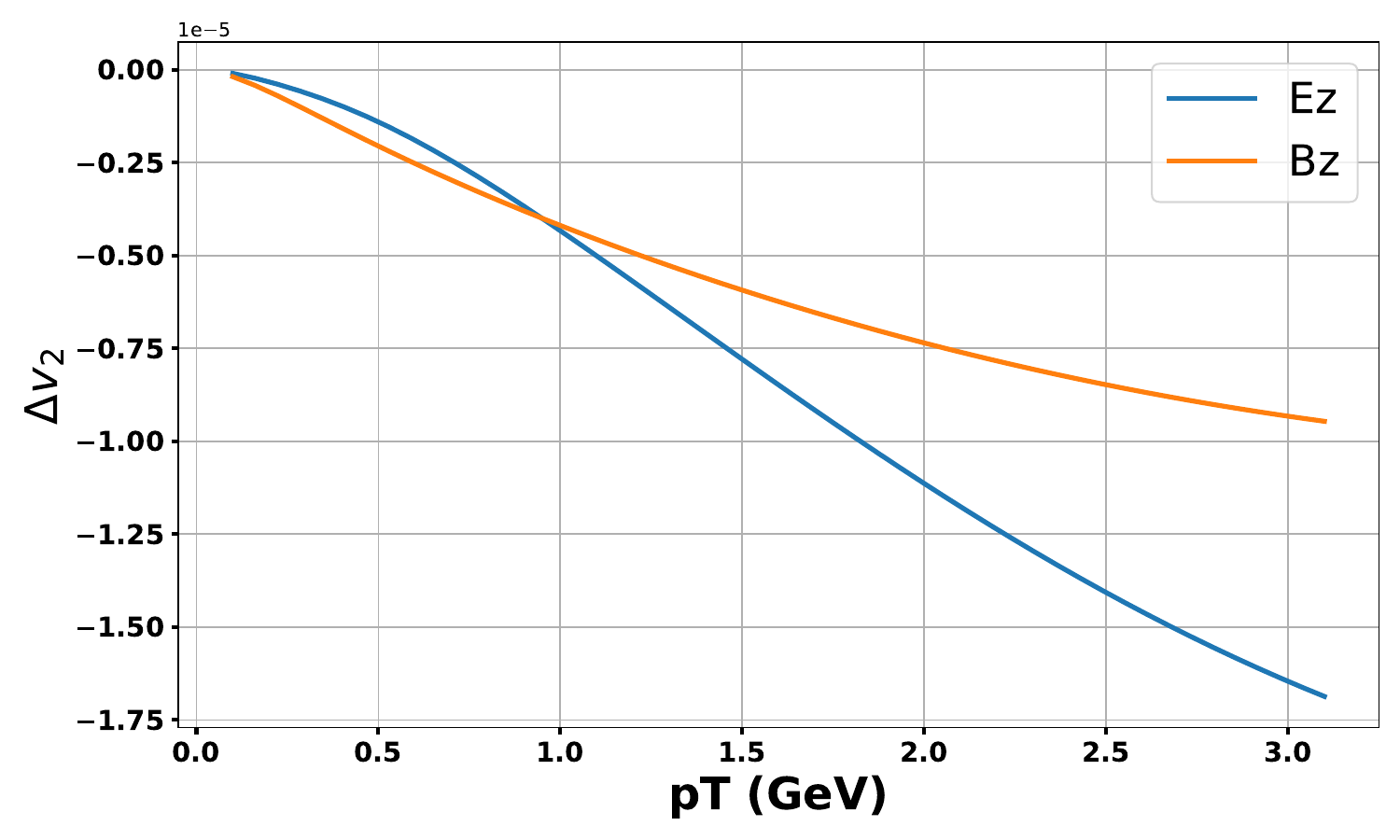}
        \caption{}
        \label{fig:rato3}
    \end{subfigure}
    \caption{(Color Online) $\Delta v_2$ vs $p_T$ plots are shown for (a) only $eE_x$ and $eE_y$, (b) only $eB_x$ and $eB_y$, and (c) only $eE_z$ and $eB_z$, respectively.
}
    \label{fig:combined}
\end{figure}

Here, we discuss the results for $\Delta v_2$, defined as $v_2^{\pi^-} - v_2^{\pi^+}$, due to the electromagnetic fields at $\sqrt{s_{\text{NN}}} = 27$ GeV. We begin by examining the individual effects of the electromagnetic field components $eF \in [eE_x, eE_y, eE_z, eB_x, eB_y, eB_z]$.
Starting with the top panel of Fig.~\eqref{fig:combined}, we present the effects of $eE_x$ and $eE_y$, depicted in solid blue and orange lines, respectively. The middle panel shows the impact of $eB_x$ and $eB_y$, while the lower panel illustrates the effects of the longitudinal field components $eE_z$ and $eB_z$, also in solid blue and orange lines, respectively.
We observe that $eE_x$ and $eB_x$ enhance $\Delta v_2$ for pions that means this leads to an enhancement in the elliptic flow for $\pi^{-}$ as compared to that of $\pi^{+}$, whereas the other components cause a decrease in the splitting with $p_T$. The enhancement due to $eE_x$ is particularly significant, while $eB_x$ shows a relatively minor increase. On the other hand, $eE_y$ exhibits a dominant effect, with a noticeable decrease comparable to the increase from $eE_x$. The longitudinal fields, however, show negligible change, likely because $v_2$, which primarily measures anisotropy in the transverse plane, is less influenced by longitudinal fields.
Moreover, the effects of the transverse magnetic fields are of the order of $10^{-5}$, relatively small compared to the electric fields' impact. This difference may be attributed to the slower decay of electric fields compared to magnetic fields, a trend clearly visible in Fig.~\eqref{fig:all_fields}.

Further insights into these effects can be gained by studying the velocity difference ($V_{\pi}$) between particles and antiparticles on the hypersurface. This difference is given as follows:
\begin{align*}
    V_{\pi} &= V_{\pi}^{-} - V_{\pi}^{+} \\
    V_{\pi}^{+} &= \sqrt{(V_x^{+})^2 + (V_y^{+})^2} \\
    V_{\pi}^{-} &= \sqrt{(V_x^{-})^2 + (V_y^{-})^2}
\end{align*}

\begin{figure*}
    \centering
    \begin{subfigure}[t]{0.35\linewidth}
        \centering
        %\raisebox{0.01\baselineskip}[0pt][0pt]{\hspace*{0.8\linewidth}\scalebox{1.0}{$1 \times 10^{-6}$}}
        \includegraphics[width=\linewidth, trim=0 0 0 0, clip]{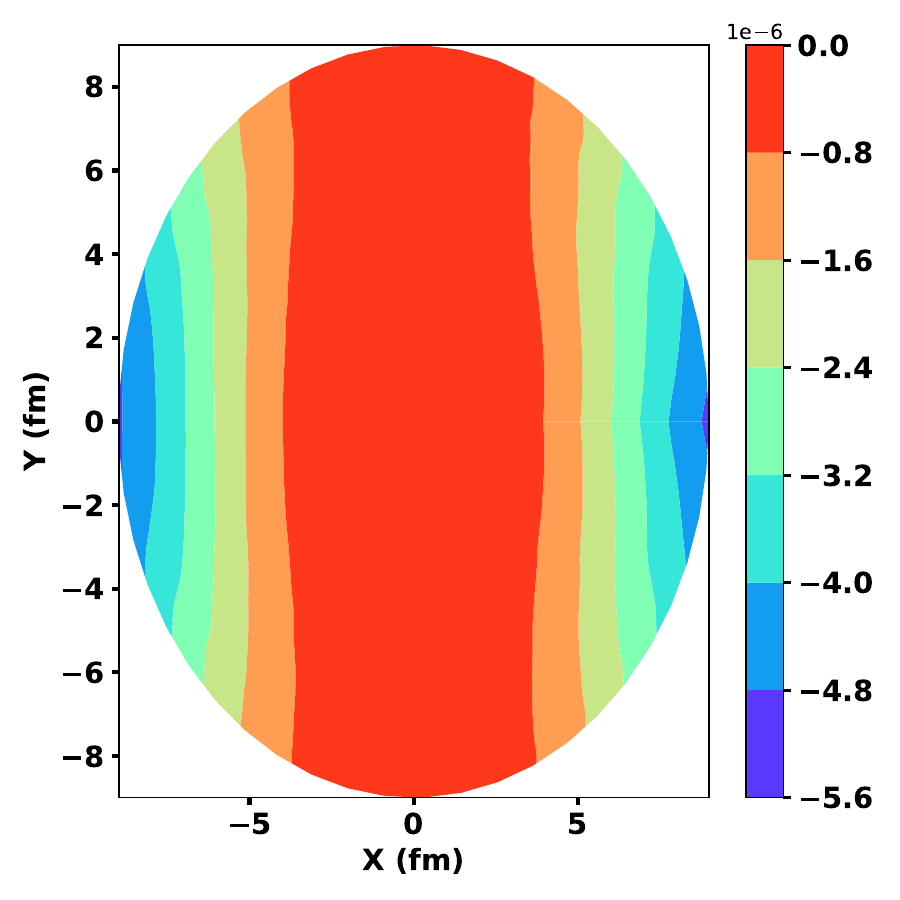}
        \caption{}
        \label{conezbz}
    \end{subfigure}
    \hspace{0.02\linewidth}
    %\hfill
    \begin{subfigure}[t]{0.35\linewidth}
        \centering
        \includegraphics[width=\linewidth, trim=0 0 0 0, clip]{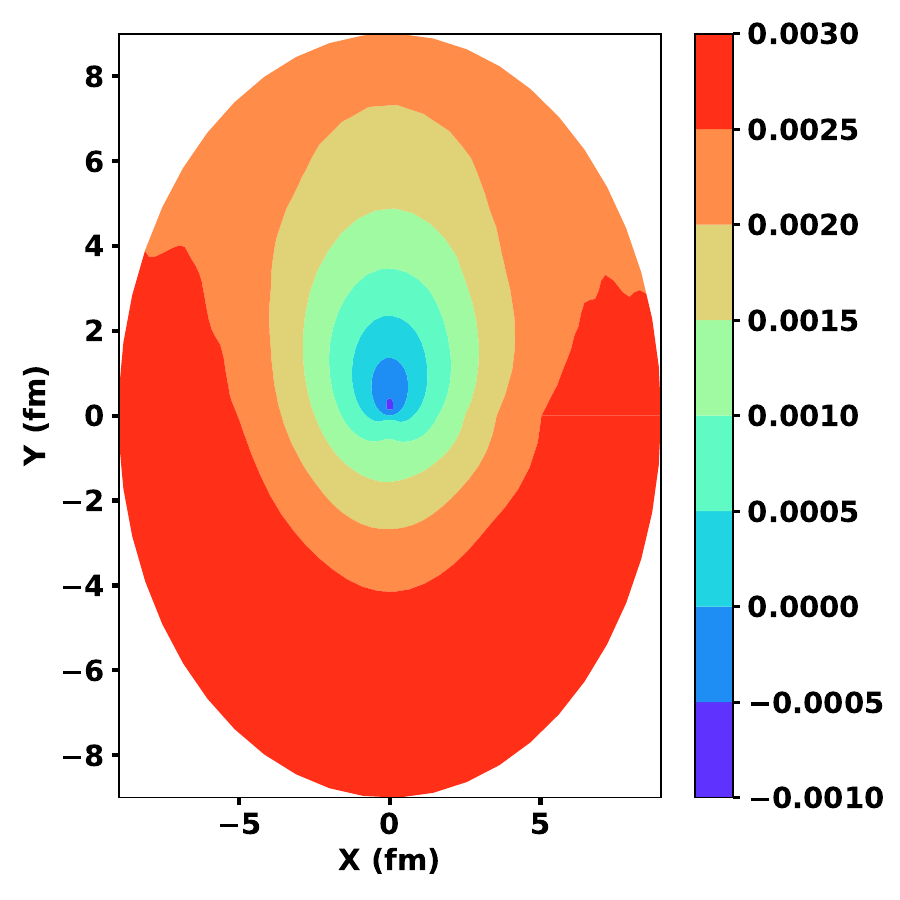}
        \caption{}
        \label{ExEy}
    \end{subfigure}
    \vspace{0.3cm}
    
    \begin{subfigure}[t]{0.35\linewidth}
        \centering
        %\raisebox{0.01\baselineskip}[0pt][0pt]{\hspace*{0.8\linewidth}\scalebox{1.0}{$1 \times 10^{-6}$}}
        \includegraphics[width=\linewidth, trim=0 0 0 0, clip]{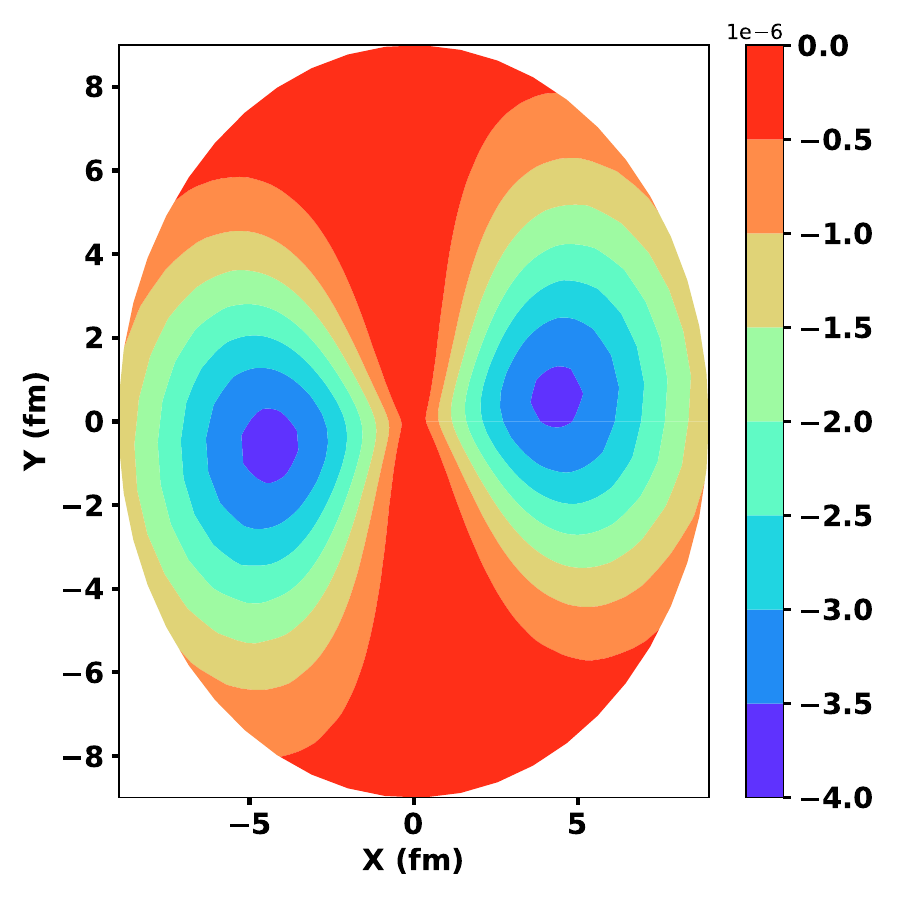}
        \caption{}
        \label{BxBy}
    \end{subfigure}
    \hspace{0.02\linewidth}
    %\hfill
    \begin{subfigure}[t]{0.35\linewidth}
        \centering
        \includegraphics[width=\linewidth, trim=0 0 0 0, clip]{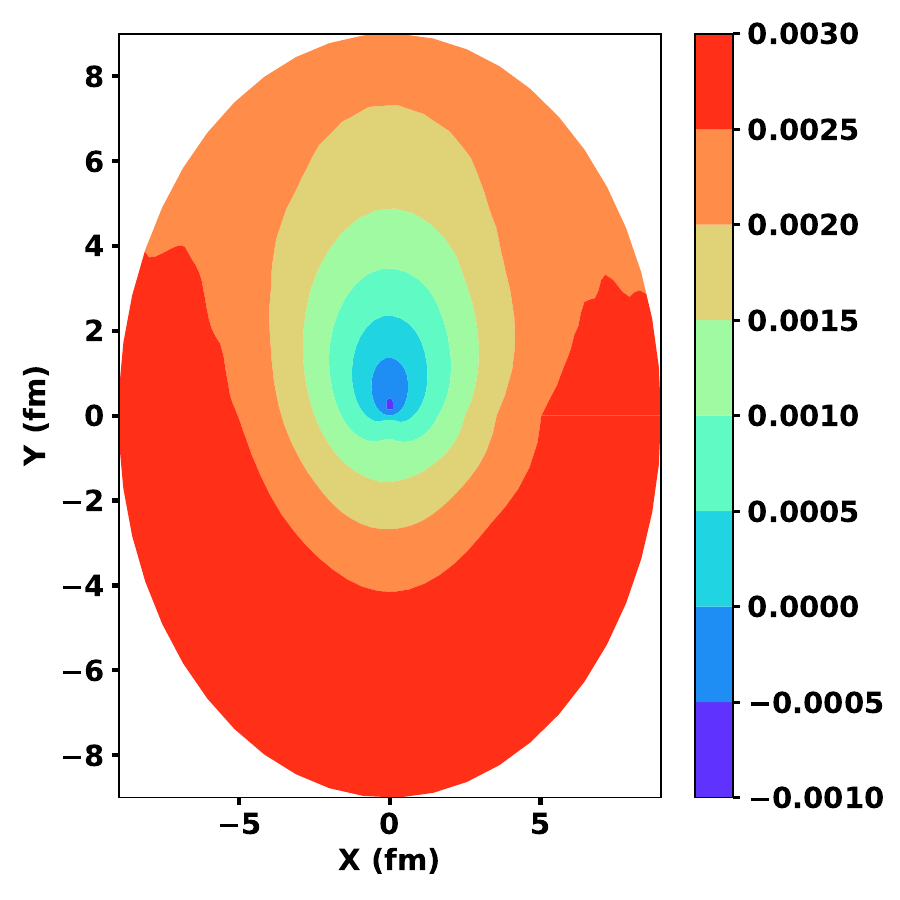}
        \caption{}
        \label{all}
    \end{subfigure}
    \caption{(Color Online) Plot for net velocity \( V_{\pi} \) on the hypersurface, where 
    (a) showcases the contribution of both \( eE_z \) and \( eB_z \) combined, 
    (b) shows the contributions from both \( eE_x \) and \( eE_y \), 
    (c) shows the contributions from both \( eB_x \) and \( eB_y \), 
    and (d) shows the contributions from all the field components combined together, respectively.}
    \label{fig:combined1}
\end{figure*}

\begin{figure}
    \centering
    \includegraphics[width=1\linewidth]{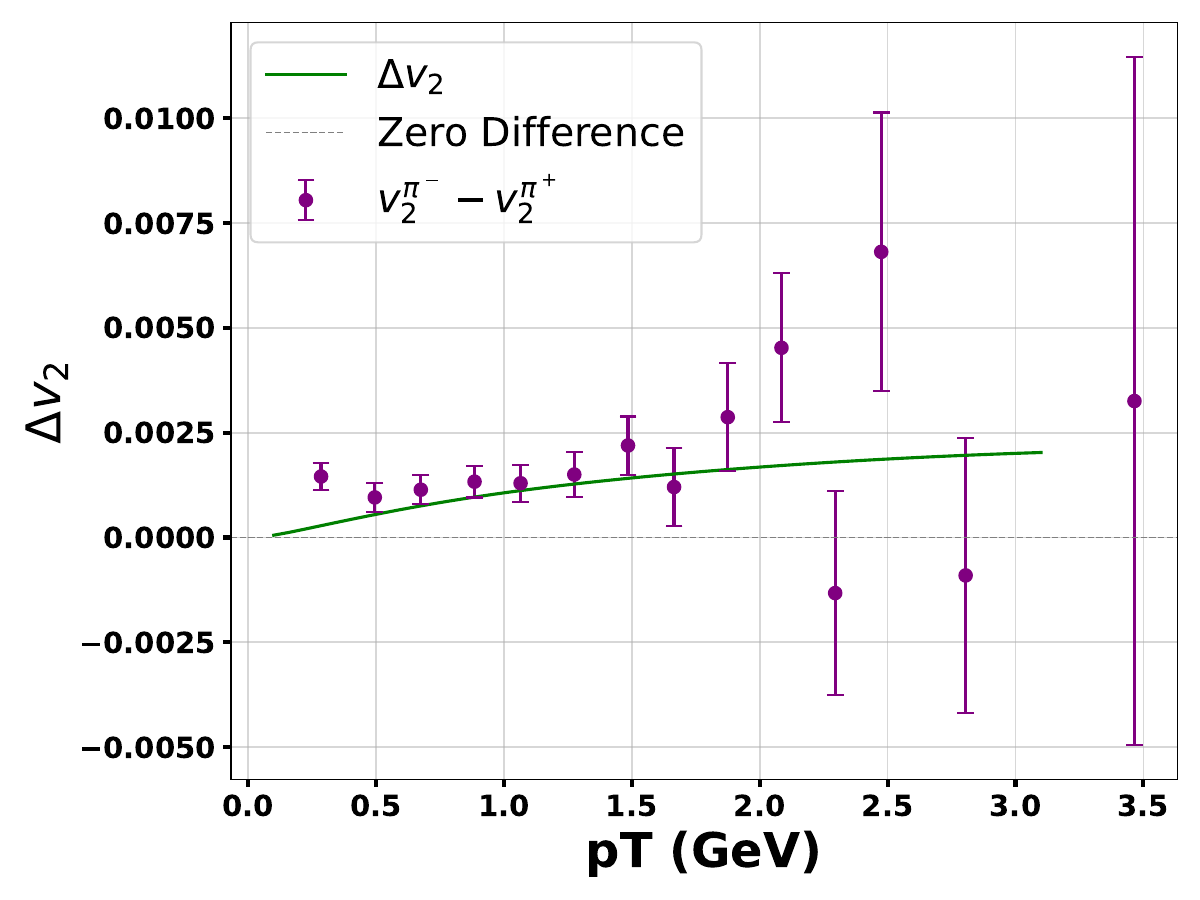}
    \caption{(Color Online) Plot of \(\Delta v_2\) as a function of \(p_T\), showcasing our fit result (color green), with the data shown with error bars taking the fields at b = 3 fm.
}
    \label{fit}
\end{figure}

In this study, the term "difference" refers to the disparity between anti-pions and pions, with an increase in $\Delta v_2$ indicating a greater flow of anti-pions relative to pions.
 Further analyzing the difference in velocities on the hypersurface, we observe the combined effects of electric and magnetic fields on the velocity splitting. As was previously mentioned the bottom panel of Fig.~\eqref{fig:combined} shows that the longitudinal components, $eE_z$ and $eB_z$, slightly suppress $\Delta v_2$, this can be clearly attributed from Fig.~\eqref{conezbz}, where the mostly negative region on the hypersurface leads to a negative $\Delta v_2$ when integrated over all the spatial points. Conversely, Fig.~\eqref{ExEy} depicts an overall enhancement due to a stronger $eE_x$, resulting in a positive $\Delta v_2$ vs $p_T$.
A smaller suppression in Fig.~\eqref{BxBy} is observed due to the dominance of $eB_y$ over the minor enhancement by $eB_x$. Finally, Fig.~\eqref{all} shows the combined effect of all field components, similar to the transverse electric field's influence, highlighting the electric field's dominance on the transverse plane in elliptic flow splitting. Further, our calculated $\Delta v_2$ is compared with the STAR experiment data~\cite{STAR:2013ayu}, showing good agreement within uncertainties, where electric fields on the transverse plane are mainly responsible for the observable splitting. This comparison is performed by slightly adjusting the $eE_x$ and $eE_y$ components, as they are highly sensitive to the drift velocity, which results in the splitting of the elliptic flow. These adjustments allow us to estimate the maximum strength of the fields that could describe the data well. The modifications in the $eE_x$ and $eE_y$ components are however minimal, kept under 10\% in each case, to achieve a reasonable fit with the experimental data. This can be seen in Fig.~\eqref{fit}.

\begin{figure}[H]
    \centering
    \includegraphics[width=1\linewidth]{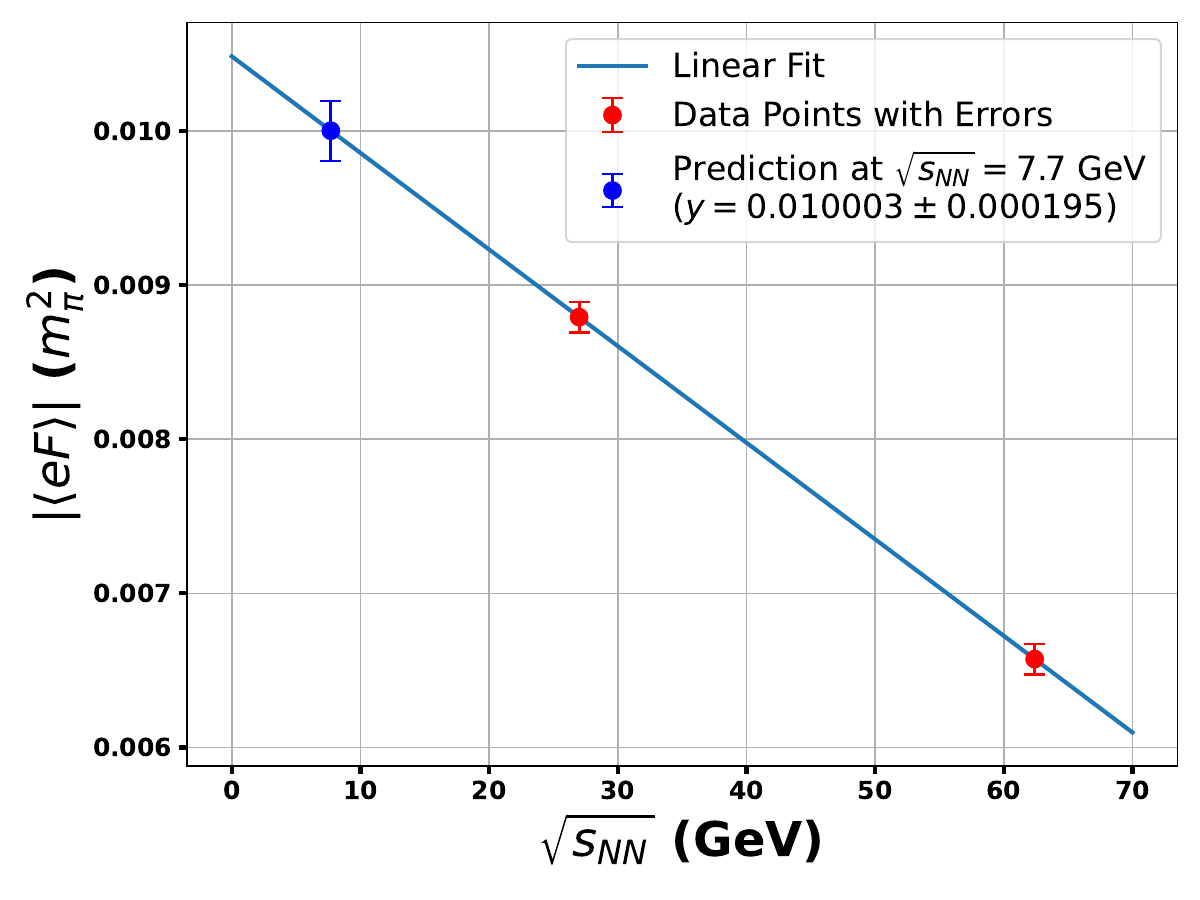}
    \caption{(Color Online) Linear fit to extract the maximum strength of the fields on the hypersurface. The red color represents the results obtained using our current formalism (depicted as data points with error bars), while the blue point with error bar correspond to the predicted result at \(\sqrt{s_{\text{NN}}} = 7.7\) GeV, as given by the formula in Eq.~\eqref{eFfit}.
 }
    \label{fitall}
\end{figure}

Previously, we also studied the effects of electric fields on $\Delta v_2$ with $p_T$ at mid-rapidity by exploring several scenarios of electric field configurations. Although the field strengths in those scenarios were unrealistic, the results qualitatively showed a similar trend in the splitting of the elliptic flow~\cite{Panda:2023akn}.
The primary finding from that study was that the relative difference in the $x$ and $y$ components of the electric field led to the observable splitting in $v_2$. Coming back to the study here, in addition to studying the effect of individual electromagnetic field contributions, we also measured a quantity defined as $|\langle eF \rangle|$, which represents the absolute value of the mean strength of all field components on the hypersurface. The mean here is taken over all spatial points on the hypersurface and across all field components.
We also calculated this quantity for a relatively higher energy, specifically at $\sqrt{s_{\text{NN}}} = 62.4$ GeV, and found that its value decreases with increasing collision energy. This decrease can be attributed as the principal reason for the smaller splitting observed at higher energies. We should also note that with increasing impact parameter, the strength of the electric fields decreases, but the strength of the magnetic fields increases~\cite{Siddique:2022ozg}. 
To determine the maximum field strength on the hypersurface, we increased the magnitude of all magnetic field components by one order, accounting for the highest achievable strength of the magnetic fields on the hypersurface.
 By doing this, we obtained the value of $|\langle eF \rangle|$ at two different $\sqrt{s_{\text{NN}}}$, specifically at 27 and 62.4 GeV.
The trend of $|\langle eF \rangle|$ with $\sqrt{s_{\text{NN}}}$ is illustrated in Fig.~\eqref{fitall}, with error bars. These error bars are obtained by calculating the variation in mean field strengths that can describe the experimental data well within the reported uncertainties from the experiments.

By using these two data points, we obtained the value of $|\langle eF \rangle|$ at $\sqrt{s_{\text{NN}}} = 7.7$ GeV using a linear fit of the following form:
\begin{eqnarray}\label{eFfit}
    |\langle eF \rangle| = \text{A} \sqrt{s_{\text{NN}}} + \text{B},
\end{eqnarray}
where $\text{A}$ and $\text{B}$ are the fit parameters obtained with certain uncertainties. This approach helps in predicting the maximum value of $|\langle eF \rangle|$ at $\sqrt{s_{\text{NN}}} = 7.7$ GeV, which comes out to be \( (0.010003 \pm 0.000195) \, m_{\pi}^2 \). The error in the predicted result is obtained using the propagation of error from the extracted fit errors of the slope and intercept values. 

This value, in principle, provides an upper limit of the mean field strengths that can be measured on the freeze-out hypersurface, which could describe this bulk observable well. However, a more rigorous approach to explaining the data would involve incorporating the complete effects of electromagnetic fields along with fluid evolution and their interactions using a comprehensive MHD code. This current study, however, serves a way to understand the individual contributions and provides a heuristic understanding of the reasons behind the splitting.

\section{Summary and conclusions}\label{conclusion}
To summarize, the present study primarily investigates the effects of all components of the electromagnetic field on the splitting of the elliptic flow in low-energy heavy-ion collisions. This work is based on a set of simplifying assumptions. The first assumption concerns the fluid velocity profile, which is modeled using the blast wave model, while the electromagnetic fields are computed in a realistic setting of heavy-ion collisions. The evolution of these fields is governed by the solutions of Maxwell's equations, under the assumption that the chiral conductivity is much smaller than the electrical conductivity, both of which are treated as constant throughout the analysis. After setting the profile for both fluid and fields, the Cooper-Frye prescription is then employed to evaluate the splitting of the elliptic flow, incorporating the effects of the electromagnetic fields by perturbatively adding the drift velocity (which accounts for these fields) to the previously parameterized fluid velocity. For this analysis, only pions are considered, as they serve as a cleaner probe for the study of the electromagnetic field effects compared to kaons and protons, where other conserved charges like strangeness and baryon chemical potential could also influence the outcome.

Our results show that the difference in the velocity distribution on the transverse plane of the hypersurface is predominantly positive for the case of \( eE_x \), while \( eB_x \) has a mild effect in enhancing this difference. Other components, however, tend to suppress the distribution, with \( eE_y \) exhibiting the most significant effect. This difference in the velocity distribution mainly from the transverse components of electric fields translates into the final splitting of the elliptic flow, which is primarily driven by the difference between in-plane and out-of-plane contributions to particle production. Consequently, we conclude that in this scenario, electric fields are the dominant contributor to the splitting.
When comparing our results with data at \( \sqrt{s_{\text{NN}}} = 27 \, \text{GeV} \), we find a good agreement within the uncertainty, which enables us to further calculate the maximum value of the average field strength \( |\langle eF \rangle| \) that would lead to such a splitting. For \( \sqrt{s_{\text{NN}}} = 27 \, \text{GeV} \) and \( 62.4 \, \text{GeV} \), we observe that this field strength increases as the center-of-mass energy decreases. This increase might explain the rise in the splitting at lower \( \sqrt{s_{\text{NN}}} \) values. Using linear interpolation, we predict the maximum value of \( |\langle eF \rangle| \) at \( \sqrt{s_{\text{NN}}} = 7.7 \, \text{GeV} \) to be \( (0.010003  \pm 0.000195) \, m_{\pi}^2 \).
 
As mentioned earlier, the present study provides a simplistic estimation of the splitting of the elliptic flow. A more rigorous and sophisticated approach could involve the use of a realistic 2+1 or 3+1 dimensional hydrodynamic code~\cite{Schenke:2010nt,Pang:2012he}, incorporating electromagnetic field effects, or the application of a state-of-the-art magnetohydrodynamic (MHD) code to gain a deeper understanding of the phenomenon. Additionally, one could explore other analyses, such as employing a tilted initial condition as an input to the hydrodynamic evolution and electromagnetic field estimation, to further refine the reason for the splitting of the flow harmonics. 
The current study can also be extended to kaons and protons, keeping in mind that additional conserved charges, such as strangeness and baryon chemical potential, could play a significant role in these cases.

% \section*{Pearson Correlation Coefficient}

% The Pearson correlation coefficient (\(r\)) is a statistical measure that quantifies the strength and direction of a linear relationship between two variables \(X\) and \(Y\). It is defined by the formula:
% \[
% r = \frac{\sum_{i=1}^{n} (X_i - \bar{X})(Y_i - \bar{Y})}{\sqrt{\sum_{i=1}^{n} (X_i - \bar{X})^2 \sum_{i=1}^{n} (Y_i - \bar{Y})^2}}
% \]

% Where:
% \begin{itemize}
%     \item \(X_i\) and \(Y_i\): Individual data points of the variables \(X\) and \(Y\).
%     \item \(\bar{X}\) and \(\bar{Y}\): Means of \(X\) and \(Y\), respectively.
%     \item \(n\): Number of data points.
% \end{itemize}

% \subsection*{Interpretation}
% \begin{itemize}
%     \item \(r = +1\): Perfect positive linear correlation. As \(X\) increases, \(Y\) increases proportionally.
%     \item \(r = -1\): Perfect negative linear correlation. As \(X\) increases, \(Y\) decreases proportionally.
%     \item \(r = 0\): No linear correlation between \(X\) and \(Y\).
%     \item \(|r|\): The absolute value of \(r\) indicates the strength of the correlation, where values closer to 1 represent stronger relationships.
% \end{itemize}

\begin{acknowledgments}
The author acknowledges Dr. Santosh Kumar Das for his kind hospitality at IIT Goa, as well as the financial support provided through his project (Project No: 2021/EMR/SKD/026). The author would also like to express sincere thanks to Dr. Victor Roy for the initial interaction.
\end{acknowledgments}

%\newpage
\bibliography{ref}

\end{document}